\documentclass[aps,pra]{revtex4}

\def\dbarit {{\mathchar'26\mkern-11mud}}
\usepackage{graphicx}
\usepackage{hyperref}
\begin{document}

\title{Quantum Heat Engines, the Second Law and Maxwell's Daemon}

\author{Tien D. Kieu}
\email[]{kieu@swin.edu.au}

\affiliation{ARC Centre of Excellence for Quantum-Atom Optics,\\
Centre for Atom Optics and Ultrafast Spectroscopy,\\
Swinburne University of Technology, Hawthorn 3122, Australia}

% \date{\today}

\begin{abstract}
We introduce a class of quantum heat engines which consists of
two-energy-eigenstate systems, the simplest of quantum mechanical
systems, undergoing quantum adiabatic processes and energy exchanges
with heat baths, respectively, at different stages of a cycle. Armed
with this class of heat engines and some interpretation of heat
transferred and work performed at the quantum level, we are able to
clarify some important aspects of the second law of thermodynamics.
In particular, it is not sufficient to have the heat source hotter
than the sink, but there must be a minimum temperature difference
between the hotter source and the cooler sink before any work can be
extracted through the engines. The size of this minimum temperature
difference is dictated by that of the energy gaps of the quantum
engines involved.  Our new quantum heat engines also offer a
practical way, as an alternative to Szilard's engine, to physically
realise Maxwell's daemon. Inspired and motivated by the Rabi
oscillations, we further introduce some modifications to the quantum
heat engines with single-mode cavities in order to, while respecting
the second law, extract more work from the heat baths than is
otherwise possible in thermal equilibria. Some of the results above
are also generalisable to quantum heat engines of an infinite number
of energy levels including 1-D simple harmonic oscillators and 1-D
infinite square wells.
\end{abstract}
%\pacs{}
%\keywords{}

\maketitle
\section{\label{sec:secondlaw}Introduction}
The second law has started out as a ``no-go" statement against a
certain class of perpetual machines but is now a pillar of modern
physics, supported by experimental evidence without exception so
far. While the first law of thermodynamics is a statement of
quantity about energy conservation, the second law is a statement of
quality about what kinds of energy transformations are and are not
allowed. T,here are several classical statements of the second
law~\cite{zemansky, callen}:
\begin{itemize}
\item {\it\bf Kelvin-Planck:}  No process is possible whose \underline{sole}
result is the absorption of heat from a reservoir and the conversion
of this heat into work.

\item {\it\bf Clausius:}  No process is possible whose \underline{sole} result
is the transfer of heat from a cooler to a hotter body.

\item {\it\bf Entropy Maximum Postulate:} The entropy of a closed system never
decreases in any process.
\end{itemize}
The first two statements above can be shown to be equivalent
by the introduction of intermediate heat engines.  And the last postulate
requires the introduction of entropy which is a function of extensive parameters
of a composite system, defined for all equilibrium states and having certain
properties.

Entropy of a composite system in a macro-state can be linked, in statistical
physics, to the statistics of the many micro-states
that correspond to the same macro-state.  In order to satisfy the statistical
nature of entropy and to derive the principle of increasing entropy,
several fundamental assumptions are required~\cite{Planck:98}:
\begin{itemize}
\item The composite system has a large number of components (atoms, molecules, etc.).
\item These components can be divided into a small number of classes of
indistinguishable components.  (In fact, were every molecule of a
gas assumed to be different from each other then statistical physics
would become fairly simple but useless as it would not be able to
account for the non-decreasing flow of entropy.)
\item Boltzmann's fundamental hypothesis~\cite{Pauli:00b}: all micro-states are
equally probable.  This is termed as {\em elementary disorder} by
Planck and has been generalised to the quantum domain as the
hypotheses of equal {\em a priori} probabilities and random {\em a
priori} phases for the quantum states of a system~\cite{Tolman:79}.
All of these can be subsumed by the (somewhat stronger) {\em ergodic
hypothesis}~\cite{Pauli:00b} which postulates that the dynamical
time average is equal to the ensemble average (of appropriate
ensembles), except for a number of exceptional initial conditions of
relatively vanishing importance.
\end{itemize}
No matter how reasonable the last assumption may be, it should be
pointed out that those hypotheses are not a part of but are extra to
the first principles of quantum mechanics.

The second law was indeed the motivation and the philosophical
reason for Max Planck to introduce the concept of energy quanta in
his solution for the puzzle of black-body
radiation~\cite{Planck:98}. His line of reasoning can be rephrased
as follows: the second law is about irreversibility in nature;
irreversibility does nothing but defines a preference of the more
probable over the less probable; and the probability assignment for
physical states (in order to have the more and the less probables)
requires those states to belong to a {\em distinguishable} variety
of possibilities.  This led Planck to the conclusion that the {\em
atomic hypothesis} must be necessary.  And the rest was history; the
concept of quanta was then born as the states of discrete
homogeneous elements.  In that context, the study of heat engines in
the quantum domain in relation to the second law is a contribution
to a completion of the circle that was started at the beginning of
the last century.

Present technology now allows for the probing and/or realisation of
quantum mechanical systems of mesoscopic and even macroscopic sizes
(like those of superconductors, Bose-Einstein condensates, etc.)
which can also be restricted to a relatively small number of energy
states.  It is thus important to study these quantum systems
directly in relation to the second law. Our study, started
with~\cite{Kieu:04} and further expanded in this paper, is part of a
growing body of investigations into quantum heat
engines~\cite{Zurek:03, Geva:1992, Feldmann:1996, Feldmann:2000,
Bender:00, Opatrny:02, Scully:03, Arnaud:03, entangled, CPSun}.
Explicitly, the only principles we will need are those of the
Schr\"odinger equation, the Born probability interpretation of the
wavefunctions and the von Neumann measurement
postulate~\cite{vonNeumann:55}.  In particular, we will not exclude,
but will make full use of, any exceptional initial conditions, as
long as they are realisable physically. However, without a better
understanding of the emergence of classicality from quantum
mechanics, we will also have to assume the thermal equilibrium Gibbs
distributions {\em for the heat baths} that are coupled to the
quantum systems. This assumption is related to the fundamental
assumptions mentioned above and is extra to those of quantum
mechanics.  Even though we do not impose this extra assumption on
the quantum mechanical systems, the steady-state distributions {\it
for the systems} will eventually reach the Gibbs distributions in
time because of the coupling with the heat baths, see
eq.~(\ref{at1}) below.

% So it is interesting to study these systems in relation to the emperical Law.
% We will not assume anything beyond those principles of QM (which includes
% the Born probability interpretation and the von Neumann measurement postulate [von Neumann]).
% That is, with regard to the above assumptions, we don't need the above hypotheses.

In order to introduce a class of heat engines operating entirely in
the framework of quantum mechanics, we will need a quantum
interpretation of the transfer of heat and performance of work in
the next Section. This interpretation is also necessary for a review
of the second law and for a possible realisation, as an alternative
to Szilard's engine, of Maxwell's daemon in Sections~\ref{sec:qhe}
and~\ref{sec:maxwell}.  The quantum heat engines are then considered
next both in thermal equilibrium and also in thermal non-steady
states, in Sections~\ref{sec:steady} and~\ref{sec:transition}.  We
find from these quantum considerations that even though the second
law is not violated in a broad sense, it needs some refinements and
clarifications.  We also demonstrate that more work can be extracted
by the engines in non-steady states than otherwise is possible in
thermal equilibrium, Section~\ref{sec:max}. Section~\ref{sec:illus}
provides an explicit numerical illustration of such capability. We
then discuss Maxwell's daemon further in Section~\ref{sec:finer} as
the reason behind any violation of the second law were we ever able
to control the quantum phases of the heat baths.
Sections~\ref{sec:sho} and~\ref{sec:inf} contain some generalised
results for quantum heat engines with simple harmonic oscillators
and infinite square wells, respectively, all in one dimension.
Finally, we end the paper with some concluding remarks in the final
Section~\ref{sec:disc}.

\section{\label{sec:heatwork}Quantum identifications of heat exchanged and work
performed}
The expectation value of the measured energy of a quantum system with
discrete energy levels is
\begin{eqnarray}
U = \langle E \,\rangle &=& \sum_i p_i\,E_i,
\end{eqnarray}
in which $E_i$ are the energy levels and $p_i$ are the corresponding
occupation probabilities.  Infinitesimally,
\begin{eqnarray}
dU &=& \sum_i\{E_i\,dp_i + p_i\, dE_i\},
\label{first}
\end{eqnarray}
from which we make the following identifications for
infinitesimal heat transferred
$\dbarit Q$ and work done $\dbarit W$
\begin{eqnarray}
\dbarit Q &:=& \sum_i E_i\, dp_i,\nonumber \\
\dbarit W &:=& \sum_i p_i\, dE_i.
\label{id}
\end{eqnarray}
Mathematically speaking, these are not total differentials but are
path dependent. These expressions interpret heat transferred to or
from a quantum system as the change in the occupation probabilities
but not in the change of the energy eigenvalues themselves; and work
done on or by a quantum system as a redistribution of the energy
eigenvalues but not of the occupation probabilities of each energy
level. Together with these identifications, equation~(\ref{first})
can be seen as just an expression of the first law of
thermodynamics, $dU = \dbarit Q + \dbarit W$.

The above link between the
infinitesimal heat transferred to the infinitesimal change of occupation
probabilities
is in accord with, or at least is not in contradiction to, the thermodynamic
link between heat and entropy,
$\dbarit Q = T\,dS$, in combination with the statistical physical link between
entropy and probabilities, $S = -k\sum_i p_i\,\ln p_i$.
On the other hand, expression~(\ref{id}) linking work performed
to the change in energy levels
agrees with the fact that work done on or by a system can only be
performed through a change
in the generalised coordinates of the system, which in turn gives rise to a
change in the distribution of the energy levels~\cite{Schrodinger:89}.

\section{\label{sec:qhe}A class of quantum heat engines}
The quantum heat engines considered herein are just two-energy-level
quantum systems, the simplest of quantum mechanical systems,
operated in a cyclic fashion described below.   (They are the
quantum analogue of the classical Otto engines and are readily
extendable to systems of many discrete energy levels.)  They could
perhaps be realised with coherent macroscopic quantum systems like,
for instance, a Bose-Einstein condensate confined to the bottom two
energy levels of a trapping potential. The exact cyclicity will be
enforced to ensure that upon completing each cycle all the output
products of the engines are clearly displayed without any hidden
effect.

\begin{figure}
\begin{center}
\includegraphics*[scale=0.4]{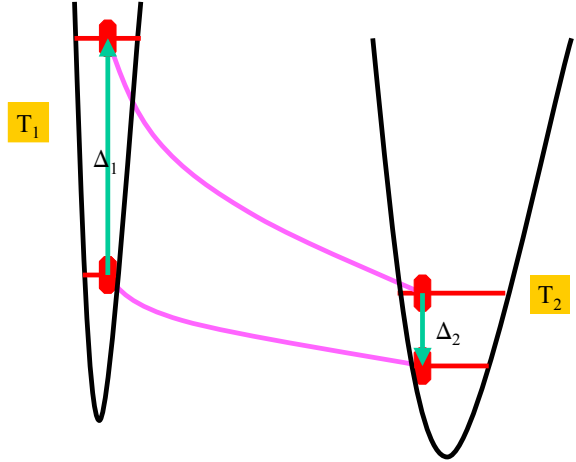}%
\caption{\label{figa}On the left hand side, a two-state quantum system
in the lower state comes into contact with a heat bath at
temperature $T_1$ for some time until it absorbs
an amount of energy $\Delta_1$ to jump into the upper state.  Next in the
passage to
the right hand side, the system undergoes a quantum adiabatic process, remaining
in the upper state, and performs work
on the relaxing potential wall.  On the right hand side, the system comes into
contact with
another heat bath at temperature $T_2$ for some time until it releases an amount
of energy $\Delta_2$ to jump back to the lower
state.  In the final passage to the left hand side to complete a cycle of the
heat engine,
the system undergoes another quantum adiabatic process in which it remains in
the lower
state and work is done on the system by the compressing potential wall.}
\end{center}
\end{figure}

A cycle of the quantum heat engine consists of four stages:
\begin{itemize}
\item {\em Stage 1:}  The system has some probability to be in the
lower state {\em prior} to some kind of contact (whose nature will
be discussed later on) with a heat bath at temperature $T_1$.  After
some time interval, there is a probability that the system receives
some energy from the heat bath to jump up an energy gap of
$\Delta_1$ to be in the upper state.  According to the
identification above, only heat is transferred in this stage to
yield a change in the occupation probabilities, and no work done as
there is no change in the values of the energy levels.  This stage
is depicted on the left hand side of Fig.~\ref{figa}.
\item {\em Stage 2:}  The system is then isolated from the heat bath and
undergoes a quantum adiabatic expansion, whose nett result is to
reduce the energy gap from $\Delta_1$ to a smaller value $\Delta_2$.
In this stage, provided the expansion rate is sufficiently slow
according to the quantum adiabatic theorem~\cite{Messiah}, the
occupation probabilities for the two states remain unchanged.  The
system may perform an amount of work.  This is depicted as the upper
branch of Fig.~\ref{figa} running from left to right. Note that
there is no change in probability so there is no heat transferred;
that is, a quantum adiabatic process implies a thermodynamic
adiabatic process (but not the other way around, in general).
\item {\em Stage 3:}  The system is next brought into some
kind of contact with another heat bath at temperature $T_2$ for some
time.  There is a probability that it releases some energy to the
bath to jump down the gap $\Delta_2$ to be in the lower state.  This
is depicted on the right hand side of Fig.~\ref{figa}.  Some heat is
thus transferred but no work is performed in this stage.
\item {\em Stage 4:}  The system is removed from the
heat bath nd undergoes a quantum adiabatic contraction, whose nett
result is to increase the energy gap from $\Delta_2$ back to the
previously larger value $\Delta_1$.  This is depicted as the lower
branch of Fig.~\ref{figa} running from right back to left.  In this
stage an amount of work is done on the system.
\end{itemize}
Ideal quantum adiabatic processes are employed here because they
yield, on the one hand, the maximum amount of work performable {\em
by} the systems in stage 2 (as the transition probabilities to the
lower state in that stage can be made vanishingly small according to
the quantum adiabatic theorem), but yield the minimum amount of work
performable {\em on} the systems (by some external agents) in stage
4, on the other hand. In each cycle the amount of work done {\em by}
the system is $(\Delta_1 - \Delta_2)$, which is also the nett amount
of heat it absorbs. Note that we need not and have not assigned any
temperature to the quantum system; all the temperatures are
properties of the heat baths, which in turn are assumed to be in the
Gibbs state.

However, in the operation above, the absorption and release of energy in
stages 1 and 3 occur neither definitely nor deterministically.
Quantum mechanics tells us that they can only happen probabilistically; and
the probabilities that such transitions take place depend on the details of the
interactions with and some intrinsic properties (namely, the temperatures) of
the heat baths.

Let ${p}_u^{\,(1,2)}$ be the probabilities for the system to be in
the upper level at the beginning of stages 2 and 4, respectively.
The nett work done by our quantum heat engines in the two quantum
adiabatic passages in stages 2 and 4 is
\begin{eqnarray}
\Delta W &=& \left(\int_{\mbox{\small left $\to$ right}} +
\int_{\mbox{\small right $\to$ left}}\right)
\sum_i p_i\, dE_i,\nonumber\\
&&\nonumber\\
&=&
\left({p}_u^{\,(1)}-{p}_u^{\,(2)}\right)\left(\Delta_2-\Delta_1\right).
\label{5}
\end{eqnarray}
We will make full use of this simple expression in the Sections
below.

By the weak law of large numbers, probability reflects the relative
occurrence frequency of an event in a large number of repetitions.
In the case of thermal equilibrium with only one single heat bath
($T_1=T_2$), even with a small probability
$\tilde{p}_u^{\,(1)}(1-\tilde{p}_u^{\,(2)})$ the system could be in
the upper level in stage 2 {\em and} in the lower level in stage 4,
where $\tilde{p}$ is the thermal equilibrium probabilities
in~(\ref{thermal}) below. The probability, however, diminishes
exponentially for $n_c$ consecutive cycles, $(\tilde{p}_u^{\,(1)}
(1-\tilde{p}_u^{\,(2)}))^{n_c}$, in all of which the system would
perform nett work on the environment. As a result, there {\em are}
certain cycles {\em whose \underline{sole} result is the absorption
of heat from a reservoir and the conversion of this heat into work,
of the amount $(\Delta_1-\Delta_2)$}!

This amounts to a violation of the Kelvin-Planck statement of the
second law due to the explicit probabilistic nature of quantum
mechanical processes. This violation of classical statements for the
second law, however, occurs only randomly, with some vanishingly
small probability {\em in the longer term}, and thus is not
controllable -- neither harnessible nor exploitable.

This may seem non-surprising given the statistical nature of the
second law, but it is somewhat different from the usual scenario of
violation of the second law by statistical fluctuations in the bulk.
One example of this latter scenario is the instantaneous
concentration of all the gas molecules in a big room into one of its
corners. Mathematically, this configuration is permissible due to
the existence of the Poincar\'e cycles in mechanics, but statistical
physics effectively rules this out (gives it a vanishing
probability) by invoking extra assumptions and hypotheses as
mentioned in Section~\ref{sec:secondlaw}. The scenario with our
quantum heat engines is different in that it only involves a single
(macro) quantum system with physically realisable quantum mechanical
probabilities, and without any extra hypothesis.  The subtle
difference between the two cases is in the time average for a single
system, on the one hand, and the bulk average for systems having
many subcomponents, on the other.

\section{\label{sec:maxwell}A realisation of Maxwell's Daemon}
However, there exists a sure way to {\em always} extract work, to the amount of
$(\Delta_1 - \Delta_2)$,
in each completable ``cycle" described in Section~\ref{sec:qhe}.  In order to
eliminate the
probabilistic uncertainty in the thermalising contacts with heat baths,
\begin{itemize}
\item we prepare the system to be in the lower state prior to stage 1;
\item we perform an energy measurement after stage 1 and then only let the engine
continue to stage 2 subject to the condition that the measurement result confirms that the system is in the
upper state; if this is not the case, we go back to the last step above;
\item we next perform another measurement after stage 3 and then only let the engine
continue
to stage 4 subject to the condition that the measurement result confirms that the system is in the lower
state; if this is not the case, we
go back to the last step above.
\end{itemize}
All the steps above is to ensure ${p}_u^{\,(1)}=1$ and
${p}_u^{\,(2)}=0$, {\em irrespective of the heat bath temperatures},
and thus to be {\em always} able to derive maximum work according
to~(\ref{5}). That is, all this can be carried out {\it even for the
case} $T_1\le T_2$ to always extract, in a controllable manner, some
work which would have been otherwise prohibited by the second law.

This apparent violation of the second law is analogous to that of
Szilard's one-atom engine and is nothing but the result of an act of
Maxwell's daemon~\cite{Demon:90}. Indeed, the condition of strict
cyclicity of each engine's cycle is broken here.  After each cycle
the measurement apparatus, being a Maxwell's daemon, has already
registered the conditioning results which are needed to determine
the next steps of the engine's operation. In this way, there are
extra effects and changes to the register/memory of the apparatus,
even if we assume that the quantum measurement steps themselves cost
no energy and leave no nett effect anywhere else. To remove these
remnants in order to restore the strict cyclicity, we would need to
bring the register back to its initial condition by erasing any
information obtained in each of the cycles, either by resetting its
bits or by thermalising the register with some heat bath. Either
way, extra effects are inevitable, namely an amount of heat of at
least $kT\ln 2$ will be released per bit erased.  This is the
Landauer principle which saves the second law. More extensive
discussions and debates on these issues can be found in the
literature~\cite{Demon:90}, and in particular in the quantum version
of Szilard's engine by Zurek~\cite{Zurek:03}. Lloyd
in~\cite{Lloyd:97} has analysed carefully and in detailed the
Landauer principle behind the working of Maxwell's daemon entirely
in the framework of quantum mechanics. Maxwell's daemon has also
been discussed~\cite{Nielsen:98} in the context of quantum error
correction in quantum computation.

Our quantum heat engines could thus provide, in a different way to
Szilard's engine, a feasible and quantum mechanical way to realise
Maxwell's daemon.

\section{\label{sec:steady}Thermal steady states}
We have pointed out in Section ~\ref{sec:qhe} that there are random
instances where our quantum heat engines can extract heat from a
single heat bath and totally turn that into work or, equivalently,
can transfer heat from a cold to a hot source (and may also do some
work at the same time).  Nonetheless, on the average, there is no
violation of the second law.

If the system is allowed to thermalise with the heat baths in stages
1 and 3, the thermal equilibrium probabilities for the two energy
levels only depend on the relevant temperatures and the energy gaps,
but not on the initial states when the system is brought into
contact with the heat bath, see~(\ref{at1}) below. More explicitly,
the probability $\tilde{p}_u^{\,(1)}$ to have the system {\it in the
upper state} at the end of stage 1 after being thermalised with the
heat bath at temperature $T_1$,  and $\tilde{p}_u^{\,(2)}$ at the
end of stage 3 after being thermalised with the heat bath at
temperature $T_2$, respectively, are
\begin{eqnarray}
% \tilde{p}_u^{\,(1)} &=& 1/\left(1 + \exp\{{\Delta_1}/{kT_1}\}\right),  \nonumber\\
\tilde{p}_u^{\,(i)} &=& 1/\left(1 +
\exp\{{\Delta_i}/{kT_i}\}\right), \mbox{ for } i = 1, 2.
\label{thermal}
\end{eqnarray}
These probabilities are definitely {\em non-zero} and bounded by
$0.5$.  From which, the work derived from~(\ref{5}) is
\begin{eqnarray}
\Delta W_{\rm th} &=&
\left(\tilde{p}_u^{\,(1)}-\tilde{p}_u^{\,(2)}\right)\left(\Delta_2-\Delta_1\right).
\label{5a}
\end{eqnarray}

This is negative (that is, work has been performed {\em by} our
engines),
 given that $\Delta_1>\Delta_2$, {\em if and only if}, as can be seen from~(\ref{thermal}),
 % $\tilde{p}_u^{\,(1)}>\tilde{p}_u^{\,(2)}$.  That is,
\begin{eqnarray}
{T_1} &>& T_2\left(\frac{\Delta_1}{\Delta_2}\right),
\label{6}
\end{eqnarray}
which is to be compared with the necessary condition of the
classical statement of the second law that $T_1$ is simply greater
than $T_2$.

One might think that the last result is just another equivalent
restatement of the second law by arguing that the entropy decreased
in the heat bath in stage 1 must be, according to the usual
statement of the second law, less than the entropy increased in the
other heat bath in stage 3,
\begin{eqnarray}
\frac{\Delta Q_2}{T_2} &>& \frac{\Delta Q_1}{T_1}\, , \label{scnd}
\end{eqnarray}
and by {\em assuming} that $\Delta Q_i = \Delta_i$, for both $i$,
upon which~(\ref{6}) would have immediately followed.  However, this
assumption is not justifiable because the heat  $\Delta Q_1$
released by the hot heat bath, {\em on the average}, cannot be the
same as the energy gap $\Delta_1$ of the system at that point; and
likewise for the heat absorbed by the colder reservoir.  These heat
amounts must be, on the average, less than the corresponding energy
gaps because the heat absorbed/released by the quantum system must
be moderated by the change in occupation probabilities
(see~(\ref{id})), but such a change in the occupation probability
for a specific level is always less than one. In other words, the
energy gaps $\Delta_i$'s are the maximum energy transfers possible
in any exchange but the probability distributions in thermal
equilibrium will not allow those maximum values to be reached.
However, the result~(\ref{6}) is {\em consistent} with the second
law~(\ref{scnd}) in the sense that it can be derived
from~(\ref{scnd}) if $\Delta Q_i$ is proportional to $\Delta_i$ and
if the proportionality constants (which are less than one) are the
same for both $i=1,\, 2$. But such a proportionality is extra
ingredient to the second law~(\ref{scnd}), and is a consequence of
the same cause that also leads to~(\ref{6}). Consequently, our
derived result~(\ref{6}) is a refinement of the classical statement
of the second law, and not simply a restatement in another
equivalent form.

The expressions~(\ref{5}) and~(\ref{6}) not only confirm the broad
validity of the second law but also refine the law further in
specifying how much $T_1$ needs to be larger than $T_2$ before some
work can be extracted.  In other words, {\em work cannot be
extracted, on the average, even when $T_1$ is greater than $T_2$ but
less than $T_2(\Delta_1/\Delta_2)$}, in contradistinction to the
classical requirement that $T_1$ only needs to be larger than $T_2$.
The refinement factor $(\Delta_1/\Delta_2)$ is necessarily greater
than unity (by the requirement of energy conservation) and is
dictated by the quantum structure of the heat engines.   This result
may be extended to multi-level quantum heat engines with appropriate
energy gaps, provided the quantum energy levels involved are
discrete. (See~\cite{CPSun}, however.)  We show in
Section~\ref{sec:sho} the equivalent condition, $ {T_1} > T_2
\left({\omega_1}/{\omega_2}\right)$, for quantum simple harmonic
oscillators--where $\omega_1$ and $\omega_2$ are, respectively, the
frequencies of the oscillators in the equivalence of stages 1 and 3
above (with $\omega_1>\omega_2$).

The efficiency of the two-state engines is found to be
\begin{eqnarray}
\eta_q &=& \frac{\Delta W_{\rm th}}{Q_{\rm in}} =
\left(1-\frac{\Delta_2}{\Delta_1}\right),
\label{eff}
\end{eqnarray}
which is {\em independent} of temperatures and is the maximum
available within the law of quantum mechanics.  (A similar
expression, but through a specific context, was also obtained
in~\cite{Scovil:59}.) This expression also serves as the upper
bound, with appropriate $\Delta_1$ and $\Delta_2$, of the efficiency
of any quantum heat engine because the work performed by our heat
engines through their quantum adiabatic processes is the maximum
that can be extracted.

The efficiency is, as a consequence of~(\ref{6}), less than that of
the classical Carnot engines, $\eta_C$,
\begin{eqnarray}
\eta_q < 1 - T_2/T_1 \equiv \eta_C.
\end{eqnarray}
This is in agreement with a general fact established by
Lloyd~\cite{Lloyd:97} that quantum efficiency must be reduced as
more information is obtained about the system either by measurement
or decoherence.  Lloyd argues that when the system is in a fully
measured or decohered state (that is, when the density matrix is
already diagonal with respect to a measured or preferred basis) then
no further information can be introduced, whence the Carnot
efficiency might thus be achieved.

\begin{figure}
\begin{center}
\includegraphics*[scale=1]{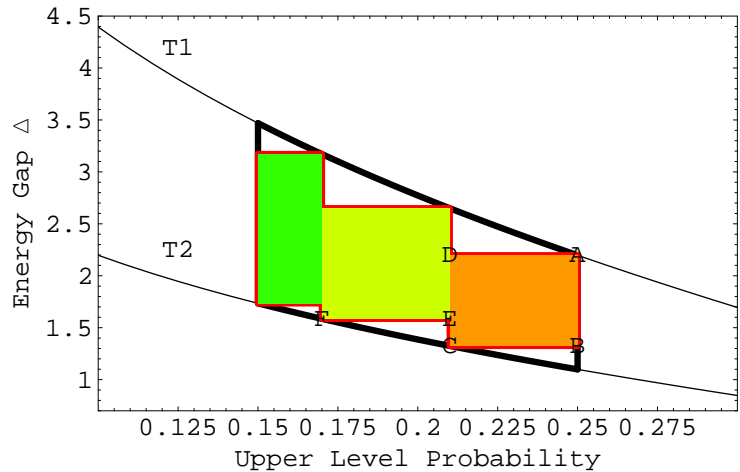}%
\caption{\label{efficiency}Even though the quantum efficiency is
always bounded by the Carnot efficiency, a quantum heat engine with
asymptotically infinite number of quantum adiabatic steps can have
both its efficiency and work output per cycle approaching those of
the latter. This can be seen when the area of the (red) zigzag
polygon approaches, from the inside, that of the irregular shape
bounded by the two outermost vertical lines and the curves labeled
$T_1$ and $T_2$. See text for further explanation.}
\end{center}
\end{figure}

However, the limiting Carnot efficiency may also be approached in
the quantum mechanical framework through the limit of an infinite
number of quantum adiabatic processes~\cite{Geva:1992, Bender:00,
Arnaud:03}. We illustrate this fact for our quantum heat engines in
Fig.~\ref{efficiency}.  A similar discussion can also be found with
the heat engines of~\cite{Arnaud:03}.

Fig.~\ref{efficiency} depicts the inverses of~(\ref{thermal}),
\begin{eqnarray}
\Delta &=& kT\ln\left(\frac{1}{\tilde{p}}-1\right),
\end{eqnarray}
with the labels $T_1$ and $T_2$ correspond to the two temperatures.
The rectangle $ABCD$ represents a particular operation of our
quantum heat engines between two heat baths $T_1$ and $T_2$ -- with
stage 2 represented by the segment $AB$ (which is adiabatic with no
change in the probability); stage 3 by $BC$ (no work done with
constant energy gap); stage 4 by $CD$; and stage 1 by $DA$,
completing an engine cycle. The area of $ABCD$, by virture
of~(\ref{5}), represents the work derivable from this operation,
which is less than the area bounded by the two vertical lines $AB$
and $CD$ and the two curves, which in turn is the work derivable
from a corresponding Carnot engine also operated between the two
temperatures.

Now, we modify our heat engines to have $AB$ as the adiabatic
expansion, $BC$ the heat exchange, $CE$ the adiabatic compression,
$EF$ the heat exchange, and so on.  When the division becomes finer
and finer with more and more steps, the area of the (red) zigzag
polygon will approach, from the inside, that of the irregular shape
bounded by the outermost two vertical lines and two horizontal
curves. This is the limit when our modified quantum heat engine can
have both the same work output per cycle and the same efficiency as
those of a corresponding Carnot engine.

We think that this approach will have some interesting consequences
in the context of quantum information and hope to be able to present
further analysis on this limiting scenario elsewhere.

\section{\label{sec:transition}Transition to thermal steady states}
Given that the efficiency is bounded by that of Carnot engines,
which is another manifestation of the second law, can we derive more
work (with a larger heat input so that the efficiency bound is
maintained) than that obtainable from the thermal steady states? In
the transient states approaching the thermal equilibria in stages 1
and 3 of a quantum heat engines' cycle, the density matrix elements
with the upper eigenstate $|u\rangle$ and lower $|l\rangle$ satisfy
the following equations~\cite{Scully:97}, for $i=1, 2$,
\begin{eqnarray}
\partial_t\rho_{uu}^{\,(i)} &=& -(\bar{n}_i +1 )\Gamma \rho_{uu}^{\,(i)}
+ \bar{n}_i\Gamma \rho_{ll}^{\,(i)},\nonumber\\
\partial_t\rho_{ll}^{\,(i)} &=& -\bar{n}_i\Gamma \rho_{ll}^{\,(i)}
+ (\bar{n}_i+1)\Gamma \rho_{uu}^{\,(i)},\\
\partial_t\rho_{ul}^{\,(i)} &=& \partial_t\left(\rho_{lu}^{\,(i)}\right)^*
= -\left( \bar{n}_i + \frac{1}{2}\right)\Gamma \rho_{ul}^{\,(i)},\nonumber
\label{decay}
\end{eqnarray}
under the Markovian assumption and the rotating-wave approximation
and in which the heat bath is treated as a collection of infinite
number of simple harmonic oscillators.  In the above, $\Gamma$ is
the decay rate and we have assumed that the thermal average boson
number in the heat bath having frequency $\nu = \Delta_i/h$ is
\begin{eqnarray}
\bar{n}_i &=& \frac{1}{e^{\frac{\Delta_i}{kT_i}}-1}\, .
\label{planck}
\end{eqnarray}
The solution for the differential equations above with the heat bath at $T_i$ is
\begin{eqnarray}
\rho_{uu}^{\,(i)}(t) &=& e^{-(2\bar{n}_i +1)\Gamma t}\left( \rho_{uu}^{\,(i)}(0) -
\tilde{p}_u^{\,(i)}\right) + \tilde{p}_u^{\,(i)}.
\label{at1}
\end{eqnarray}

Let the system stay in contact with the heat bath at temperature $T_1$ in stage 1
for a time $\tau_1$ (without achieving thermalisation); and for a time $\tau_2$
at $T_2$ in stage 2.  The cyclicity of the quantum heat engines requires that
\begin{eqnarray}
\rho_{uu}^{\,(1)}(0) &=& \rho_{uu}^{\,(2)}(\tau_2), \nonumber\\
\rho_{uu}^{\,(2)}(0) &=& \rho_{uu}^{\,(1)}(\tau_1).
\end{eqnarray}
This requirement together with that of $\tilde{p}_u^{\,(1)} > \tilde{p}_u^{\,(2)}$ imply,
from~(\ref{at1}),
\begin{eqnarray}
0 < &\rho_{uu}^{\,(1)}(\tau_1) - \rho_{uu}^{\,(2)}(\tau_2) &<
\tilde{p}_u^{\,(1)} - \tilde{p}_u^{\,(2)},
\end{eqnarray}
for finite $\tau_1$ and $\tau_2$.  Subsequently, the work, $\left|\Delta W_{\rm tr}\right|$,
that can be derived from transient states at finite $\tau_1$ and $\tau_2$
is always less than that from thermal equilibrium, $\left|\Delta W_{\rm th}\right|$,
\begin{eqnarray}
\left|\Delta W_{\rm tr}\right| &=& \left(\rho_{uu}^{\,(1)}(\tau_1) - \rho_{uu}^{\,(2)}(\tau_2)
\right)\left(\Delta_1 - \Delta_2\right),\nonumber\\
&<& \left(\tilde{p}_u^{\,(1)} - \tilde{p}_u^{\,(2)}
\right)\left(\Delta_1 - \Delta_2\right),\\
&<& \left|\Delta W_{\rm th}\right|.\nonumber
\end{eqnarray}
We suspect that, as long as the assumption of Gibbs distributions is made for
the heat baths, a non-Markovian treatment or dropping the
rotating wave approximation would not change this last result.
Nonetheless, we present  in the next Section a modification of
the quantum heat engines  which can better the work derivation than that
which is maximally available from thermal equilibrium.

\section{\label{sec:max}Maximising the work extraction}
Inspired and motivated by the Rabi flopping for two-level systems,
see~\cite{Scully:97}, for example, we present in this Section a
modification of the engines such that more work than usual can be
derived from {\em thermal heat sources} (as contrast to a
single-Fock-state field that drives the Rabi flopping), but at the
same time more heat input would be needed in such a way that the
Carnot efficiency is still a valid upper bound. That is, no
violation of second law is claimed here despite of the enhanced work
output.

A scenario for maximizing the work output from our quantum heat
engines is as follows. Firstly, the system is prepared to be in the
lower state and then subject to a radiation field in a Fock state
which has exactly $n_1$ quanta with a frequency in resonance with
the energy gap $\Delta_1$. After some fixed time $\tau_1$, depending
on the system-field coupling strength and on the number $n_1$, Rabi
oscillations driven by the radiation field will bring the system to
the upper energy state with certainty. At this point the system can
be removed from the field to perform some work in an adiabatic
process which reduces the energy gap to $\Delta_2$. Then it is next
subject to another field of Fock state $|n_2\rangle$ which has a
frequency in resonance with the new gap $\Delta_2$. After some time
$\tau_2$ the system will be in the lower state with certainty; upon
which it can be decoupled for an adiabatic compression to complete a
cycle of the operation.

In effect, the steps above will remove the probability difference
factor in~(\ref{5}), ensuring that a nett work of $(\Delta_1 -
\Delta_2)$ is derived in each cycle. The key point here, however, is
that a Fock-state field is {\em not} a thermal field, and extra work
or extra information would be required to maintain the Fock state
such that the nett book keeping (when full cyclicity is strictly
enforced) will show that we cannot ultimately violate the second
law.

Let us exploit this Rabi mechanism and see how it will behave in a
thermal field.

\begin{figure}
\begin{center}
\includegraphics*[scale=0.4]{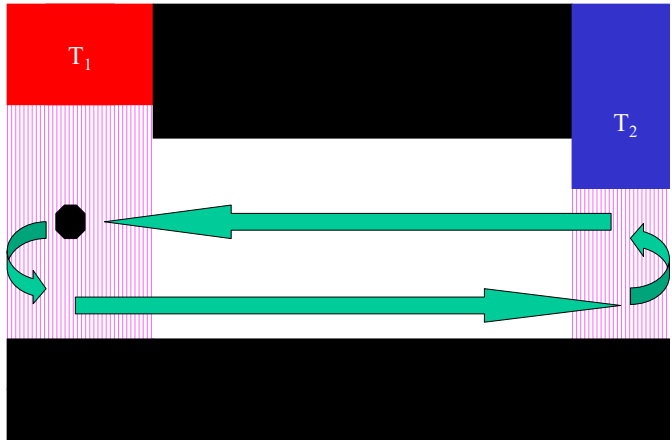}%
\caption{\label{figb}A single-mode cavity is in thermal equilibrium with a bath of
black-body radiation at temperature $T_1$.  A two-state quantum system spends some time in
the cavity whose mode matches the energy gap $\Delta_1$ between its two states.
After some prescribed time, it leaves the cavity and performs work in an quantum adiabatic
process.  It then enters another single-mode cavity which is in thermal equilibrium with
another bath of black-body radiation at temperature $T_2$ and whose mode matches the
new energy gap $\Delta_2$.  After a carefully controlled time, it leaves the cavity
and moves quantum adiabatically back to the first cavity, having work done on it to have the
energy gap increases back to $\Delta_1$.  With precise
control of the time duration spent in each cavity, the system can extract more
work in a cycle than it can if it is otherwise
let to thermally equilibrate with the two heat baths in turn. }
\end{center}
\end{figure}

A cycle of the modified quantum heat engine is depicted in
Fig.~\ref{figb}.
It also consists of four stages, of which stage 2 and stage 4 remain the same
as described in Section~\ref{sec:qhe}, whereas stages 1 and 3 are
replaced respectively by:
\begin{itemize}
\item {\em Stage 1':}  The system has a probability $p_u^{\,(1)}(0)$ to be
in its upper state.  It is entered to a single-mode cavity which is tuned to
match the energy gap $\Delta_1$ of
the system and which is in thermal equilibrium with a heat bath at temperature $T_1$.
The average occupation of the only mode survived in the cavity has thus a
(Bose-Einstein) thermal distribution $\bar n_1$
given by expression~(\ref{planck}).  After some carefully controlled time interval
$\tau_1$, the system is removed from the cavity to enter stage 2.  The probability
to find the system in it upper state is now $p_u^{\,(1)}(\tau_1)$.
Note that, as
discussed previously, only heat could be transferred in this stage
to yield a change in the occupation probabilities, and no work done as there is no change
in the values of the energy levels.  This stage is depicted on the left hand side of
Fig.~\ref{figb}.
\item {\em Stage 3':}  The system has a probability $p_u^{\,(2)}(0)$ to be
in its upper state.  It is entered into another single-mode cavity which is
in thermal equilibrium with another heat bath at temperature $T_2$ and which is tuned
to match the new energy gap $\Delta_2$ of the system.  The average occupation of the only
mode survived in the cavity has thus a (Bose-Einstein) thermal distribution $\bar n_2$
given by expression~(\ref{planck}).  After some carefully controlled time interval
$\tau_2$, the system is removed from the cavity to enter stage 4.  The probability
to find the system in it upper state is now $p_u^{\,(2)}(\tau_2)$.  This is depicted
on the right hand side of Fig.~\ref{figb}.  Some heat is thus transferred but no work is
performed in this stage.
\end{itemize}
With the quantum adiabatic processes in stage 2 and stage 4,
the cyclicity of the heat engines demands that
\begin{eqnarray}
p_u^{\,(1)}(0) &=& p_u^{\,(2)}(\tau_2), \nonumber\\
p_u^{\,(2)}(0) &=& p_u^{\,(1)}(\tau_1).
\label{initial}
\end{eqnarray}
On the other hand, the exit probability can be obtained as, with $i = 1, 2$,
\begin{eqnarray}
p_u^{\,(i)}(t) &=& \left(\mbox{initial probability in } |u_i\rangle \right) \times
\left(\mbox{transition probability from } |u_i\rangle \mbox{ to } |u_i\rangle
\mbox{ in } t\right) +\nonumber\\
&& \left(\mbox{initial probability in } |l_i\rangle  \right) \times
\left(\mbox{transition probability from }|l_i\rangle \mbox{ to } |u_i\rangle
\mbox{ in } t\right), \nonumber\\
&=& p_u^{\,(i)}(0)|\langle u_i|\psi_u^{\,(i)}(t) \rangle|^2
+ \left( 1 - p_u^{\,(i)}(0)\right)|\langle u_i|\psi_l^{\,(i)}(t) \rangle|^2.
\label{exit}
\end{eqnarray}
$|\psi_{u(l)}^{\,(i)}(t)\rangle$ is the state which starts out
in the upper (lower) state -- i.e., $|\psi_{u}^{\,(i)}(0)\rangle = |u_i\rangle$ and
$|\psi_{l}^{\,(i)}(0)\rangle = |l_i\rangle$ in the cavity at temperature $T_i$.

With the thermal distribution~(\ref{planck}) assumed for the heat baths in contact with
the cavities, the probability to find exactly $n$ photons of frequency $\Delta_i/h$
in the cavity at temperature $T_i$ is
\begin{eqnarray}
P_n(T_i) &=& \frac{1}{1+\bar{n}_i}\left(\frac{\bar{n}_i}{1+\bar{n}_i}\right)^n.
\label{thermaldist}
\end{eqnarray}
In each of the cavities so described, the state of the engines satisfies the
Schr\"odinger equation for a single two-level system interacting with a single-mode
field which has the frequency matching the engine's energy gap,
\begin{eqnarray}
i\hbar \frac{\partial |\psi^{\,(i)}\rangle}{\partial t} &=&  \hbar g\left(
\sigma_+^{\,(i)} a^{\,(i)} + \sigma_- ^{\,(i)}a^{\,(i)\dagger}\right)
|\psi^{\,(i)}\rangle, \mbox{ $i=1, 2$},
\label{Rabi}
\end{eqnarray}
with $g$ is the coupling constant between the the quantum heat engine
and the cavity mode, and the operators $\sigma_{\pm}=(\sigma_x + i\sigma_y)/2$
act on the two-state space of the engine and $a$ and $a^\dagger$ are the operators on the
Fock space of the field.  The solution of this Schr\"odinger equation is
given in~\cite{Scully:97}, from which we derive, through~(\ref{exit})
and~(\ref{thermaldist}), the probabilities
\begin{eqnarray}
p_u^{\,(i)}(t) &=& \frac{(1+2\bar{n}_i)\,p_u^{\,(i)}(0)
- \bar{n}_i}{(1+\bar{n}_i)}\sum_{n=0}^\infty P_n(T_i)
\cos^2\left(\Omega_n t\right) +
% p_u^{\,(i)}(t) &=& \left(\frac{1+2\bar{n}_i}{(1+\bar{n}_i)}\,p_u^{\,(i)}(0)
% - \frac{\bar{n}_i}{(1+\bar{n}_i)}\right)\sum_{n=0}^\infty P_n(T_i)
% \cos^2\left(\Omega_n t\right) +
\frac{\bar{n}_i( 1 - p_u^{\,(i)}(0))}{1+\bar{n}_i},\nonumber\\
&:=& A_i \sum_{n=0}^\infty P_n(T_i)\cos^2\left(\Omega_n t\right) +B_i,
\label{upperprobability}
\end{eqnarray}
where $\Omega_n = g\sqrt{n+1}$.

Note that when the sum on the rhs of the last equation collapses to
a single summand term, $P_n(T_i) \to \delta (n - n_0)$,
corresponding to the radiation field being in some Fock state
$|n_0\rangle$, then we will have the Rabi oscillation in the level
populations, which could then be exploited to apparently derive more
work than otherwise allowed by the second law, but this is only
apparent and cannot violate the law at all as discussed earlier.
% Because of the normalisation of $P_n(T_i)$,
% \begin{eqnarray}
% \left| p_u^{\,(i)}(t) - B_i \right|&\le& \left|A_i\right| \sum_{n=0}^\infty P_n(T_i)
% = \left| A_i\right|.
% \label{bounds}
% \end{eqnarray}

\begin{center}
\begin{figure}
\includegraphics[scale=0.6]{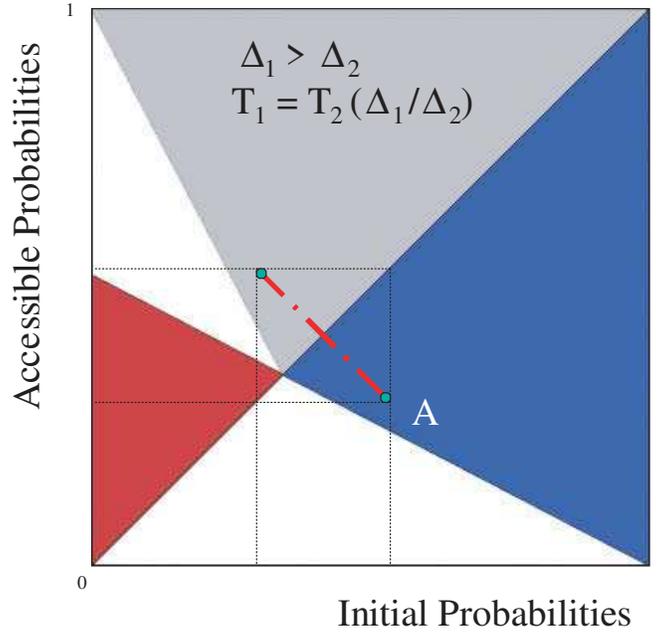}%
\caption{\label{fig1}For the cavity in contact with
a heat sink at $T_2$ where we want the initial probability to be lowered upon leaving,
the accessible leaving probability is bounded in the (blue) area under the diagonal, with
its reflection shown in gray above the diagonal.
The bounded (red) area above the diagonal is for the cavity in contact with a heat bath
at $T_1 = T_2(\Delta_1/\Delta_2)$.  The
reflection of the point A at $T_2$ across the diagonal is clearly not in the accessible
region at $T_1$, as there is no overlapping of the two regions above the
diagonal.  So we {\em cannot} form a cyclic quantum heat engine which does work
at these temperatures, {\em even when $T_1$ is greater than $T_2$ by a factor
$(\Delta_1/\Delta_2)$.}}
\end{figure}
\end{center}

From the last expression we can find the bounds, as functions of the initial probability
$p_u^{\,(i)}(0)$ and the temperature $T_i$, of the probability
$p_u^{\,(i)}(t)$ {\em for all $t$}.
Figure~\ref{fig1} depicts these bounds.  The probability
for the system to be in the upper state upon leaving a single-mode cavity in contact
with a heat bath as a function of the initial probability (upon entering the heat bath)
is only accessible in the bounded areas (of red and blue).
\begin{itemize}
\item For $p_u^{\,(i)}(0)\ge p_{\rm critical}^{\,(i)}$, where
\begin{eqnarray}
p_{\rm critical}^{\,(i)} &=& \frac{\bar{n}_i}{1+2\bar{n}_i},
\end{eqnarray}
the coefficient $A_i$ of the first term in~(\ref{upperprobability}) is positive, and so is the
first term itself.  Thus, for all $t$,
\begin{eqnarray}
B_i \le p_u^{\,(i)}(t) \le A_i \sum_{n=0}^\infty P_n(T_i)+ B_i  = A_i +B_i,
\end{eqnarray}
because of the normalisation $\sum_{n=0}^\infty P_n(T_i)=1$.  Substitution of
$A_i$ and $B_i$ yields
\begin{eqnarray}
\frac{\bar{n}_i( 1 - p_u^{\,(i)}(0))}{1+\bar{n}_i} \le p_u^{\,(i)}(t)
\le p_u^{\,(i)}(0), \, \mbox{for } p_{\rm critical}^{\,(i)} \le p_u^{\,(i)}(0) \le 1 .
\label{bound1}
\end{eqnarray}
This bounded region is depicted (in blue) in Fig.~\ref{fig1}.
\item For $p_u^{\,(i)}(0)\le p_{\rm critical}^{\,(i)}$, the coefficient $A_i$ is negative,
and so is the first term.  Thus,
for all $t$,
\begin{eqnarray}
A_i + B_i \le p_u^{\,(i)}(t) \le B_i.
\end{eqnarray}
That is,
\begin{eqnarray}
p_u^{\,(i)}(0)\le p_u^{\,(i)}(t)\le
\frac{\bar{n}_i( 1 - p_u^{\,(i)}(0))}{1+\bar{n}_i}, \,
\mbox{for } 0 \le p_u^{\,(i)}(0) \le p_{\rm critical}^{\,(i)}.
\label{bound2}
\end{eqnarray}
This bounded region is also depicted (in red) in the same Figure.
\end{itemize}
Note also that the vertex on the diagonal separating these two regions determines the
stationary point where the probability is time independent and is equal to the initial
probability.
As the thermal equilibrium probability given by the Gibbs distribution
must be independent of both time and initial probability,
it is represented by a horizontal line crossing this vertex.
For the cavity in contact with the heat bath at $T_1$,
we want to have the exit probability to be greater (the greater, the more work
can be extracted) than the initial
probability,
\begin{eqnarray}
p_u^{\,(1)}(\tau_1) &>& p_u^{\,(1)}(0) = p_u^{\,(2)}(\tau_2),
\end{eqnarray}
thus we need only to consider the appropriate portion of the bounds for this cavity,
namely that above the diagonal.
Reversely, for the cavity in contact with the heat bath at $T_2$, we want to have
the exit probability to be smaller (the smaller, the better) than the initial probability,
\begin{eqnarray}
p_u^{\,(2)}(\tau_2) &<& p_u^{\,(2)}(0) = p_u^{\,(1)}(\tau_1),
\end{eqnarray}
hence for this cavity we need only to consider the other portion of the bounds below the diagonal.
Thus, we can also use this Figure~\ref{fig1} to elucidate the situation for the two heat
baths in which $T_1 = T_2(\Delta_1/\Delta_2)$, the (red) area above the diagonal
comes from $T_1$ and the (blue) below the diagonal from $T_2$.
The coordinate of a point $A$ in the (blue) area below the diagonal in Fig.~\ref{fig1} is
$(p_u^{\,(2)}(0), p_u^{\,(2)}(\tau_2))$, representing an exit probability less than
the entry one at the cavity with temperature $T_2$.  The corresponding point at temperature
$T_1$ must have, by requirement of cyclicity, the coordinate
\begin{eqnarray}
(p_u^{\,(1)}(0), p_u^{\,(1)}(\tau_1)) &=& (p_u^{\,(2)}(\tau_2), p_u^{\,(2)}(0)),
\end{eqnarray}
which thus is the reflection of the point $A$ across the diagonal.  The gray area above the
diagonal in Fig.~\ref{fig1} is the reflection of the (blue) area for $T_2$.  However, it is clearly seen
that for $T_1 = T_2(\Delta_1/\Delta_2)$ this reflection is {\em not} in the accessible
(red) area for $T_1$.  We then conclude that at these temperatures, {\em even with $T_1$ greater than $T_2$
by a factor $(\Delta_1/\Delta_2)$}, the quantum heat engines {\em cannot} do work
{\em on the average}.  As a consequence of the temperature-independent efficiency
similar to~(\ref{eff}), the engines cannot neither absorb nor transfer any heat.

\begin{figure}
\begin{center}
\includegraphics[scale=0.6]{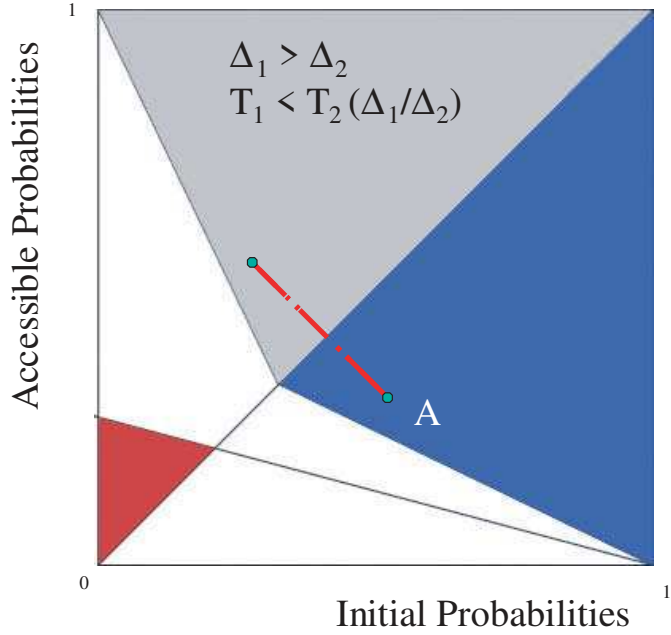}%
\caption{\label{fig2}Similar to Fig.~\ref{fig1} but this time with $T_1<
T_2 (\Delta_1/\Delta_2)$.  Once again we cannot have a cyclic quantum heat engine,
even for $T_1$ in the range $T_2<T_1<T_2 (\Delta_1/\Delta_2)$.}
\end{center}
\end{figure}

In Fig.~\ref{fig2}, we combine the relevant bounds for $T_2$ below
the diagonal (in blue) and those for $T_1$ above the diagonal (in
red) for the choice $T_1 < T_2(\Delta_1/\Delta_2)$.  It is seen once
again that no work is derivable on the average.   Not only we have
thus confirmed the second law that, {\em on the average}, no process
is possible whose \underline{sole} result is the transfer of heat
from a cooler to a hotter body, with or without a production of
work. But we have also clarified the degrees of coolness and hotness
in terms of the quantum energy gaps involved before such a process
is possible; namely, we must have $T_1 > T_2(\Delta_1/\Delta_2)$, as
in~(\ref{6}) once again.

\begin{figure}
\begin{center}
\includegraphics[scale=0.6]{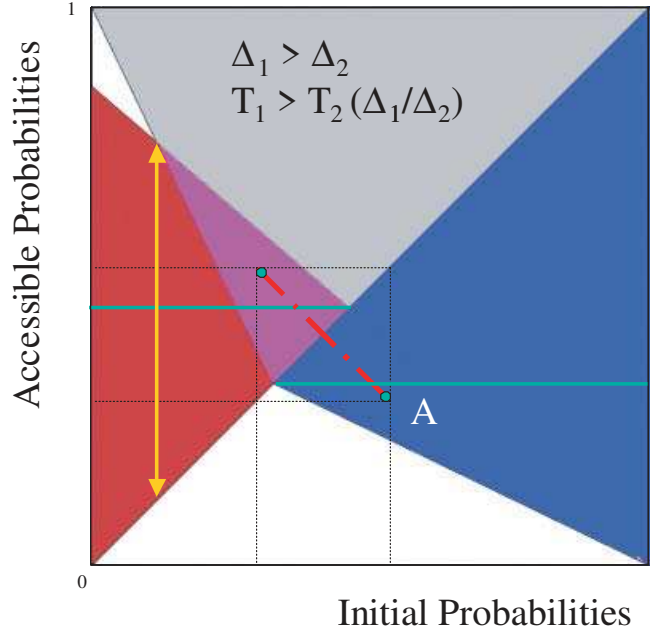}%
\caption{\label{fig3}Similar to Fig.~\ref{fig1} but this time with
$T_1> T_2(\Delta_1/\Delta_2) > T_2$.  We can form a cyclic quantum
heat engine at these temperatures, in broad agreement with the
classical statement of second law.  The maximum work that can be
extracted in a single cycle is proportional to the length of the
(yellow) vertical double arrow and is more than that for thermally
equilibrated situation, which is proportional to the distance
between the two horizontal (green) lines which represent the thermal
equilibrium probabilities for these $T_1$ and $T_2$ respectively.}
\end{center}
\end{figure}

We now show how our quantum
heat engines are capable of performing more work than can be derived from
thermal equilibrium otherwise.  In Fig.~\ref{fig3}, which combines the case
$T_1>T_2 (\Delta_1/\Delta_2)$, there is some overlap between the (red)
area for $T_1$ and the reflection of the area for $T_2$ across the diagonal.
In this case, it can be seen that the production of some work, $\Delta W_{\rm cav}$, is
now possible,
\begin{eqnarray}
\left|\Delta W_{\rm cav}\right| &=& \left(p_{u}^{\,(1)}(\tau_1) - p_{u}^{\,(2)}(\tau_2)
\right)\left(\Delta_1 - \Delta_2\right).
\label{work}
\end{eqnarray}

In general, if and when we choose to operate with a point below the thermal equilibrium
line in the (blue) area for $T_2$ such that its reflection across the diagonal is above the
thermal equilibrium line in the (red) area for $T_1$ as shown
in the figure, we can derive more work than the case of thermal equilibrium.
The work done is proportional to the difference in probabilities as shown in~(\ref{5})
and~(\ref{work}).  Here $\left|\Delta W_{\rm cav}\right|$ is greater than the
work derivable at thermal equilibrium,
$\left|\Delta W_{\rm th}\right|$,
because the vertical distance between point $A$ and its reflection in
Fig.~\ref{fig3} is greater than the vertical distance between the two horizontal lines,
which represent the two thermal equilibria.

From the bounds~(\ref{bound1}) and~(\ref{bound2}),
we can evaluate the maximum amount of work extractable
from our modified quantum heat engines at given temperatures
\begin{eqnarray}
\frac{\max \left|\Delta W_{\rm cav}\right|}{\left|\Delta W_{\rm th}\right|}
&=& \frac{\left(1+2\bar{n}_1\right)\left(1+2\bar{n}_2\right)}
{\left(1+\bar{n}_1+\bar{n}_2\right)}> 1.
\end{eqnarray}

\section{\label{sec:illus}An illustration}
\begin{figure}
\begin{center}
\includegraphics*{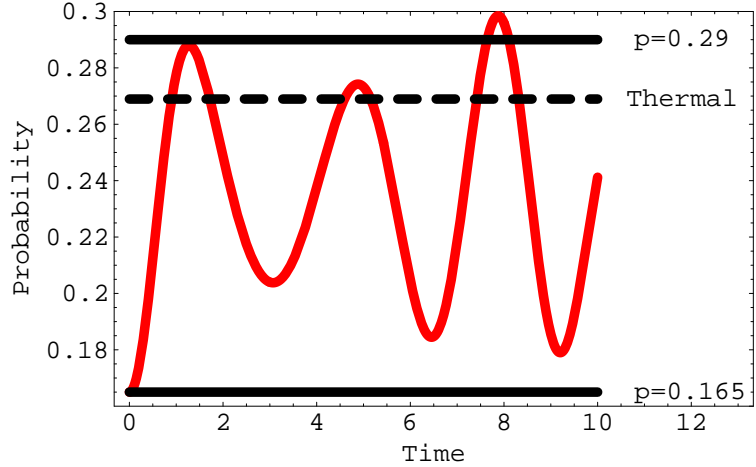}%
\caption{\label{fig4}$p_{u}^{\,(1)}$ versus time according to
eq.~(\ref{upperprobability}). The system enters a single-mode
cavity, in contact with a heat source at temperature $T_1$ (here,
$kT_1/\Delta_1 = 1.5$), with the probability in the upper state
$p_{u}^{\,(1)}(0)  = 0.165$ ($= p_{u}^{\,(2)}(\tau_2)$, by cyclicity
requirement) and leaves the cavity, at suitably chosen time
$\tau_1$, with the increased probability $p_{u}^{\,(1)}(\tau_1) =
0.29$.  This latter probability is larger than the thermal
equilibrium probability at this temperature (dashed line).}
\end{center}
\end{figure}

\begin{figure}
\begin{center}
\includegraphics*{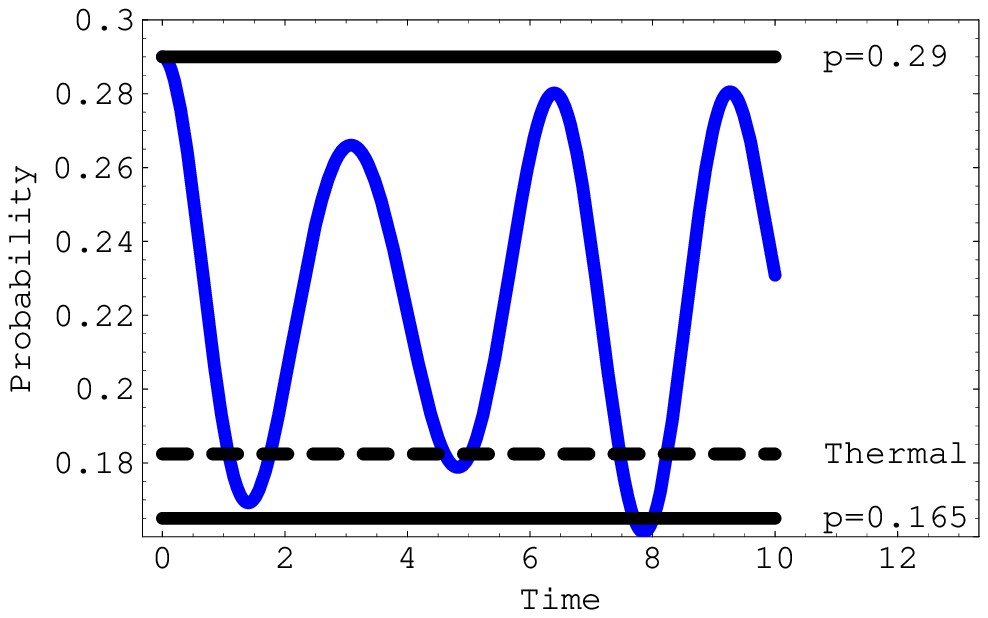}%
\caption{\label{fig5}$p_{u}^{\,(2)}$ versus time.  Similar to Fig.~\ref{fig4} but this time the probability upon
entering a cavity, in contact with a heat sink at temperature $T_2 < T_1
(\Delta_2/\Delta_1) < T_1$ (here, $kT_2/\Delta_2 = 1.0$),
is $p_{u}^{\,(2)}(0) = 0.29$ ($= p_{u}^{\,(1)}(\tau_1)$, by cyclicity requirement)
and upon leaving, at suitably
chosen time $\tau_2$, is $p_{u}^{\,(2)}(\tau_2)= 0.165$, completing a cycle of the
quantum heat engine.  The latter probability is lower than the thermal equilibrium probability
at this temperature (dashed line), thus allowing more work to be extracted.}
\end{center}
\end{figure}

As an illustration that we can choose and {\em a priori} fix the time $\tau_1$ and $\tau_2$
for all the cycles of the modified mode of operation for our quantum heat engines such
that more work can be derived than otherwise available from thermal equilibrium,
we present herein the example in Figs.~\ref{fig4} and~\ref{fig5}.  These numerical results
are obtained
from~(\ref{upperprobability}) with $g=1$ and other parameters as stated in the captions.
Note that in Fig.~\ref{fig4} the probability at subsequent time is always more than that of
the initial time, in agreement with the fact that the (red) area for $T_1$ is above the diagonal in
Fig.~\ref{fig3}.  The reverse is true for Fig.~\ref{fig5}, because the (blue) area for $T_2$ is below
the diagonal in Fig.~\ref{fig3}.

\section{\label{sec:finer}Maxwell's daemon revisited}
The expression~(\ref{upperprobability}) for the probability, derived from~(\ref{exit}),
requires some careful justifications.  Following~\cite{Scully:97}, we expand
the state vector $|\psi\rangle$ (dropping the superscript $(i)$) in terms of $|u, n\rangle$,
in which the engine is in the upper state $|u\rangle$ and
the field has exactly $n$ photons, and of $|l, n\rangle$ in which the engine is
in the lower state $|l\rangle$,
\begin{eqnarray}
|\psi\rangle &=& \sum_n \left(
c_{u,n}(t)|u, n\rangle + c_{l,n}(t)|l, n\rangle\right).
\end{eqnarray}
From this, the Schr\"odinger equation~(\ref{Rabi}) can now be replaced by
\begin{eqnarray}
\dot c_{u,n} &=& -ig\sqrt{n+1}\,c_{l,n+1},\nonumber\\
\dot c_{l,n+1} &=& -ig\sqrt{n+1}\,c_{u,n}.
\end{eqnarray}
The general solutions for these probability amplitudes are
\begin{eqnarray}
c_{u,n}(t) &=& c_{u,n}(0)\cos\left(\Omega_n t\right)
- i c_{l,n+1}(0)\sin(\Omega_n t),\nonumber\\
c_{l,n+1}(t) &=& c_{l,n+1}(0)\cos(\Omega_n t)
- i c_{u,n}(0)\sin(\Omega_n t),
\end{eqnarray}
in which the initial probability amplitudes are assumed to be factorised,
\begin{eqnarray}
c_{u,n}(0) &=& \sqrt{p_u(0)}\times\sqrt{P_n},\nonumber\\
c_{l,n+1}(0) &=& e^{i\theta_{s}}\sqrt{1-p_u(0)}\times e^{i\theta_{f}}\sqrt{P_{n+1}},
\end{eqnarray}
where the relative phases $\theta_{s}$ for the engine and $\theta_{f}$
for the field, respectively, are not zero in general.

Now, we can also express the probability on the lhs of~(\ref{upperprobability}) as
\begin{eqnarray}
p_u(t) &=& \sum_n |c_{u,n}(t)|^2.
\end{eqnarray}
Direct substitution of the amplitudes above, however, leads an additional cross
term which does not exist on the rhs of expression~(\ref{upperprobability}) but is
proportional to
\begin{eqnarray}
\sim \Re\left(i e^{i\theta_{s}}\sqrt{p_u(0)\left(1-p_u(0)\right)}\times
\sum_ne^{i\theta_{f}}\sqrt{P_n P_{n+1}}\,\cos\left(\Omega_n t\right)
\sin\left(\Omega_n t\right)\right).
\label{phase}
\end{eqnarray}
Accordingly, the various bounds for $p_u(t)$ will be modified by the term
$\sqrt{p_u(0)\left(1-p_u(0)\right)}$, which is nonlinear in $p_u(0)$, unlike the situation
in~(\ref{bound1}, \ref{bound2}) depicted in
Fig.~\ref{fig1}, from which the second law was seen to be followed.
One might think that this extra non-linear term could be exploited to beat the second law
(thanks to the newly emerged overlapping of regions which were not overlapped previously
in Fig.~\ref{fig1}).
But this is not meant to be, however.  The reason for this impossibility is that, in general,
the various phases in~(\ref{phase}) are not fixed but can be random, especially for
the phase $\theta_f$ of
the thermal field.  They will have to be averaged over, rendering the cross term~(\ref{phase})
vanished after all.  We thus get back the expression~(\ref{upperprobability}) exactly.

If the phases could be controlled physically then the second law might be violated.  But the
point to be emphasised here is that such a control of the phases will require some careful
operation which is nothing
more than just another disguised act of Maxwell's daemon.  In the end, any such resulted
violation of the second law, if possible, is not and should not be surprising at all.

\section{\label{sec:sho}Quantum heat engines with 1-D simple harmonic oscillators}
We now generalise some of the above results to quantum systems
having an infinite number of energy levels.  Firstly, we consider a
1-D simple harmonic oscillator of frequency $\omega_1$ in thermal
equilibrium with a heat bath at $T_1$.  The oscillator is then
removed from the heat bath to undergo a quantum adiabatic expansion
until its frequency is dropped to $\omega_2$.  It is then
equilibrated with another heat bath at temperature $T_2$ before
undergone another quantum adiabatic compression to raise its
frequency back to $\omega_1$.  In one such cycle, the oscillator
performs an amount of work,
\begin{eqnarray}
\Delta W &=& \sum_{n=0}^\infty \left(E_n^{\,(1)}-E_n^{\,(2)}\right)
\left(p_n^{\,(2)}-p_n^{\,(1)}\right), \nonumber\\
&=& \hbar \left(\omega_1
- \omega_2\right)\sum_{n=0}^\infty \left(n+\frac{1}{2}\right)
\left(p_n^{\,(2)}-p_n^{\,(1)}\right), \nonumber\\
&=& \hbar \left(\omega_1
- \omega_2\right)\sum_{n=0}^\infty n
\left(p_n^{\,(2)}-p_n^{\,(1)}\right),
\label{workA}
\end{eqnarray}
with the Gibbs distributions for $i = 1, 2$
\begin{eqnarray}
p_n^{\,(i)} &=& \left(1 - e^{-\alpha_i}\right)e^{-n\alpha_i},
\end{eqnarray}
and
\begin{eqnarray}
\alpha_i &=& \frac{\hbar \omega_i}{kT_i}.
\end{eqnarray}
We have used the energy expressions for simple harmonic oscillators
\begin{eqnarray}
E_n^{\,(i)} &=& \left(n+\frac{1}{2}\right)\hbar\omega_i.
\label{SHO}
\end{eqnarray}

Let
\begin{eqnarray}
x = \alpha_2 - \alpha_1
\label{x}
\end{eqnarray}
and consider the sum in~(\ref{workA}) as a
function of $x$,
\begin{eqnarray}
f(x) &=& \sum_{n=0}^\infty n \left\{
\left(1 - e^{-\alpha_1 - x}\right)e^{-n\alpha_1 -n x}
- \left(1 - e^{-\alpha_1}\right)e^{-n\alpha_1}
\right\}.
\label{f}
\end{eqnarray}
Its derivative is
\begin{eqnarray}
f'(x) &=& - \sum_{m=1}^\infty m e^{-m\left(\alpha_1+x\right)},
\end{eqnarray}
which is always negative (we always have
$x > -\alpha_1$ for the convergence of
the infinite sum, as $\alpha_2>0$).  Thus $f(x)$ is a monotonically
decreasing function; in particular, $f(x)< f(0) = 0$  for $x\ge 0$.
From the definition of $x$~(\ref{x}) and $f(x)$~(\ref{f}) we conclude that the
oscillator can only perform work on the environment ({\em i.e.} when $f(x)<0$) {\em if
and only if} $x>0$, which means that $\alpha_2 > \alpha_1$.  That is,
\begin{eqnarray}
T_1 &>& T_2 \left(\frac{\omega_1}{\omega_2}\right)
\label{2ndA}
\end{eqnarray}
is the necessary and sufficient condition for work to be performed by a
quantum simple harmonic oscillator in such a cycle described above.

\section{\label{sec:inf}Quantum heat engines with 1-D infinite square wells}
Similarly to the last Section, but we now replace the oscillator by a particle of
mass $m$ in an 1-D infinite square well.  The work performed in a cycle is also
\begin{eqnarray}
\Delta W &=& \sum_{n=0}^\infty \left(E_n^{\,(1)}-E_n^{\,(2)}\right)
\left(p_n^{\,(2)}-p_n^{\,(1)}\right),
\label{workB}
\end{eqnarray}
but this time with the energies
\begin{eqnarray}
E_n^{\,(i)} &=& n^2\frac{\hbar^2\pi^2}{2mL^2_i}
\label{energy}
\end{eqnarray}
and the thermal distributions
\begin{eqnarray}
p_n^{\,(i)} &=& \frac{e^{-\beta_i E_n^{\,(i)}}}{\sum_m e^{-\beta_i E_m^{\,(i)}}},
\end{eqnarray}
where $\beta_i = 1/(kT_i)$ and $L_i$ are the widths of the wells at $T_i$.

Let
\begin{eqnarray}
y &=&\beta_2 - \beta_1,
\label{y}
\end{eqnarray}
and consider the rhs of~(\ref{workB}) as a function $g$ of $y$.  Its derivative
wrt $y$ can be written as
\begin{eqnarray}
g'(y) &=& \langle E^{(2)}(E^{(1)}-E^{(2)})\rangle_y +
\langle E^{(1)}-E^{(2)}\rangle_y \langle E^{(2)}\rangle_y ,
\label{deriv}
\end{eqnarray}
where
\begin{eqnarray}
\langle O\rangle_y  &\equiv& \sum_n O_n p_n^{\,(2)},
\end{eqnarray}
note that $p_n^{(2)}$ is a function of $y$.

Substituting the energy expressions~(\ref{energy}) into the derivative~(\ref{deriv}),
with $L_2>L_1$,
\begin{eqnarray}
g'(y) &=& \left(\langle n^4\rangle_y  - \langle n^2\rangle_y ^2\right)
\frac{\hbar^2\pi^2}{2mL^2_2}\left(\frac{\hbar^2\pi^2}{2mL^2_2}
- \frac{\hbar^2\pi^2}{2mL^2_1}\right),
\end{eqnarray}
which is not positive, but can vanish, thanks to the well-known
positivity property of the
expression in the first pair of curly parentheses.  Noting that $g(0) = 0$, we
have $\Delta W = g(y)> g(0) = 0$ {\em if and only if} $y<0$; that is
{\em if and only if}
\begin{eqnarray}
T_1 &<& T_2.
\label{2nd}
\end{eqnarray}
This is the condition upon which the system {\em cannot} do work and
and which is in accordance
with the second law of thermodynamics.

Note also that the above mathematical derivation can also be applied
to the simple harmonic oscillators of Section~\ref{sec:sho} once we
replace the energy expressions~(\ref{energy}) by those
of~(\ref{SHO}).  However, the ability to exactly evaluate the
partition functions of the simple harmonic oscillators of
Section~\ref{sec:sho} enables us to derive the
condition~(\ref{2ndA}), which is not contradictory to but has a more
positive interpretation than the condition~(\ref{2nd}) above: even
though $\Delta W = g(y) \le 0$ {\em if and only if} $y\ge 0$, we
cannot conclude, because of the possibility that $g(y)$ can vanish,
that the system can do work if $T_1 \ge  T_2$.  This thus is
consistent with~(\ref{2ndA}), which is the condition whence work can
be extracted.

\section{\label{sec:disc}Concluding remarks}
By interpreting work and heat, but without referring to entropy
directly, in the quantum domain and by applying this interpretation
to the simplest quantum systems, we have not only confirmed the
broad validity of the second law but also been able to clarify and
refine its various aspects.  On the one hand, explicitly because of
the probabilistic nature of quantum mechanics, there do exist
physical processes which can violate certain classical statements of
the second law.  However, such violation only occurs randomly and
thus it cannot be exploitable nor harnessible.  On the other hand,
the second law is seen to be valid {\em on the average}. This
confirmation of the second law is in accordance with the fact that,
while we can treat the quantum heat engines purely and entirely as
quantum mechanical systems, we still have to assume the Gibbs
distributions for the heat baths involved. Such distributions can
only be derived~\cite{Landau:80} with non-quantum-mechanical
assumptions, which ignore, for example, any quantum entanglement
within the heat baths.  Indeed, it has been shown
that~\cite{Lenard:78, Tasaki:00} the law of entropy increase is a
mathematical consequence of the initial states being in such general
equilibrium distributions. This illustrates and highlights the
connection between the second law to the unsolved problems of
emergence of classicality, of quantum measurement and of decoherence
which are inter-related and central to quantum mechanics. Only until
some further progress is made on these problems, the classicality of
the heat baths will have to be assumed and remained hidden in the
assumption of the Maxwell-Boltzmann-Gibbs thermal equilibrium
distributions.

Even our results support the second law, on the average, we have
further clarified the degree of temperature difference~(\ref{6}), in
terms of the quantum energy gaps involved, between the heat baths
before any work can be extracted.  While the Carnot efficiency is an
upper bound for that of the quantum heat engines, the former could
be approached by the quantum engines with the introduction of an
infinite number of alternating adiabatic and heat transferred steps.
The implication of this approach in the context of quantum
information deserves further investigations elsewhere.

Inspired and motivated by the Rabi oscillations, we have also shown
how to extract more work from the heat baths than otherwise possible
with thermal equilibrium distributions -- but more heat input would
also be needed in such a way that the Carnot efficiency is still a
valid upper bound. Note that such an operation is subject to the
bounds given in~(\ref{bound1}) and~(\ref{bound2}), which then, as
can be seen through their depiction in the figures, ensure that we
stay within the second law, but refined with the necessary
condition~(\ref{6}). The perfect agreement between the specific
results derived from the quantum dynamical bounds~(\ref{bound1})
and~(\ref{bound2}) with the general result~(\ref{6}) derived from
statistical mechanics is quite remarkable. We speculate that such
agreement is not accidental but is a consequence of the Gibbs
distributions assumed for the heat baths in both derivations; and
the agreement should thus be independent of specific details of the
quantum dynamics involved. Note also that the modified operation
with single-mode cavities is not an operation of Maxwell's daemon
because the information about the time intervals $\tau_1$ and
$\tau_2$ is fixed and forms an integrated part of the modified
engines.  This information, being common to {\em all} cycles, need
not and  should not be erased after each cycle to preserve the
cyclicity condition. Other studies of very different classes of
quantum heat engines~\cite{Scully:03, entangled} have apparently
claimed similar results that more work can be derived than from
classical engines. (However, if the cyclicity condition is not
strictly observed for a heat engine, as in the case for some of
those studies, then some extra effects may be hidden or unaccounted
for, such as those associated with a maintenance of some coherent
states or some Fock state, and {\em apparent} violation of the
second law might thus be possible.)

Our class of quantum heat engines can also readily offer a feasible
way to physically realise Maxwell's daemon, in a way different to
Szilard's engine but also through the acts of quantum measurement
and information erasure.  Finally, some of our results above have
also been generalised to quantum heat engines having an infinite
number of energy levels, specifically the 1-D simple harmonic
oscillators and 1-D infinite square wells.  Some analysis of
three-level quantum heat engines has also been available
recently~\cite{CPSun}, but because of the many different energy gaps
available (similar to the case of square well), different conditions
with different combinations of gap ratios could enter a counterpart
of~(\ref{5}) above.

\begin{acknowledgments}
I would like to acknowledge Alain Aspect for his seminar at
Swinburne University which directly prompted the investigation reported herein.
I would also like to thank Bryan Dalton, Jean Dalibard, Peter Hannaford, Alan Head,
Peter Knight
and Bruce McKellar for helpful discussions; Jacques Arnaud, Michael Nielsen, Bill Wootters
and Wojciech Zurek for email correspondence.
% Peter Knight has also informed me of reference~\cite{Bender:00}.
\end{acknowledgments}

\bibliography{qhengine}

\end{document}